\documentclass[a4paper]{jpconf}
\usepackage{graphicx}
\usepackage{mediabb}
\begin{document}
\title{Final source eccentricity measured by HBT interferometry with the event shape selection}

\author{Takafumi Niida for the PHENIX Collaboration}

\address{University of Tsukuba, 1-1-1 Tennoudai, Tsukuba, Ibaraki 305-8571, Japan}

\ead{niida@bnl.gov}

\begin{abstract}
Azimuthal angle dependence of the pion source radii has been measured applying the event shape engineering technique at the PHENIX experiment.
When events with higher magnitude of second-order flow vector are selected, the oscillation of the source radii is enhanced as well as $v_2$ which leads to the enhancement of the measured final source eccentricity.
The event twist effect in the spatial source distribution in the final state has been also explored with AMPT model. Results indicate a possible twisted source due to the initial longitudinal fluctuations.
\end{abstract}

\section{Introduction}
Higher-order flow coefficients $v_n$ are useful observables to constrain the properties of the quark-gluon plasma, such as a shear viscosity over entropy density ratio, in a heavy ion collision~\cite{chvn_phnx,CGCvsVn}. 
The $v_2$ is mainly caused by the almond shape of the nuclear overlap region, 
but higher-order flow, $v_3$, $v_4$, $\cdots$, especially their odd components are originating from the initial spatial fluctuations of participant nucleons.
Recently ATLAS experiment has presented results of event-by-event $v_n$~\cite{ebeVn_atlas}, where
$v_n$ values show large variations even in the fixed centrality bin due to initial fluctuations.
To control such an initial fluctuation, the event shape engineering was suggested~\cite{ESE}.
This could be a useful tool to study the response of initial state to the system evolution and connect the initial and final states.
In these proceedings, we present results on HBT measurements using charged pions and applying the event shape engineering technique for Au+Au collisions at $\sqrt{s_{_{NN}}}$=200 GeV recorded with the PHENIX experiment.

The event shape engineering focuses on the fluctuations in the transverse plane, but the presence of fluctuations in longitudinal direction 
could cause a twisted source along that direction~\cite{Bozek}.
The number of participants going to the forward and backward directions is not necessarily the same, 
and also participant eccentricities and participant planes might be different at both angles, which leads to different event plane angles between forward and backward rapidities. Thus the initial twist may survive as a twisted flow in the final state~\cite{ETW1,ETW2}. 
In these proceedings, we examine the possibility of a spatially twisted source in the final state using HBT interferometry in AMPT model.

\section{HBT measurements with event shape selection at PHENIX}
\begin{figure}[t]
\begin{minipage}{0.5\textwidth}
\includegraphics[width=\textwidth, trim=10 45 315 580, clip]{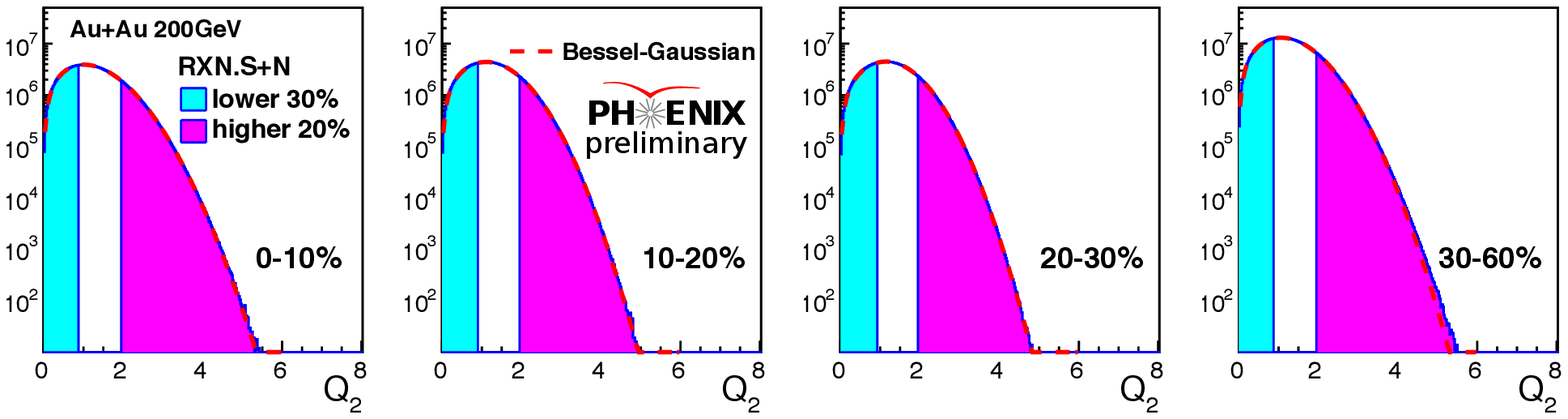} \hspace{2pc}%
\end{minipage}\hspace{2pc}%
\begin{minipage}{0.5\textwidth}
\includegraphics[width=0.90\textwidth, trim= 15 45 302.5 560, clip]{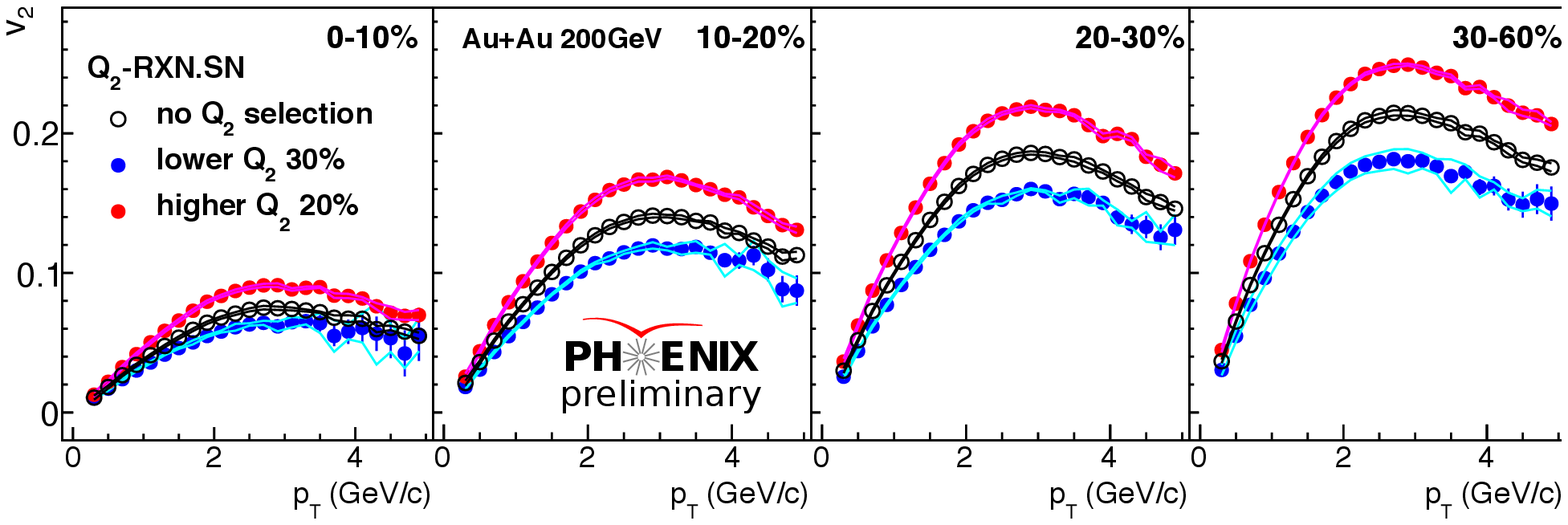}
\end{minipage} 
\caption{\label{fig1} (Left) $Q_2$ distributions measured with the RXN in Au+Au 200 GeV collisions, indicating 20\%(30\%) higher(lower) $Q_2$ events.
(Right) Charged hadron $v_2$ as a function of $p_{\rm T}$ with and without event shape selection.}
\end{figure}

\begin{figure}[tbh]
\begin{minipage}{0.6\textwidth}
\includegraphics[width=\textwidth, clip,trim=100 40 110 560]{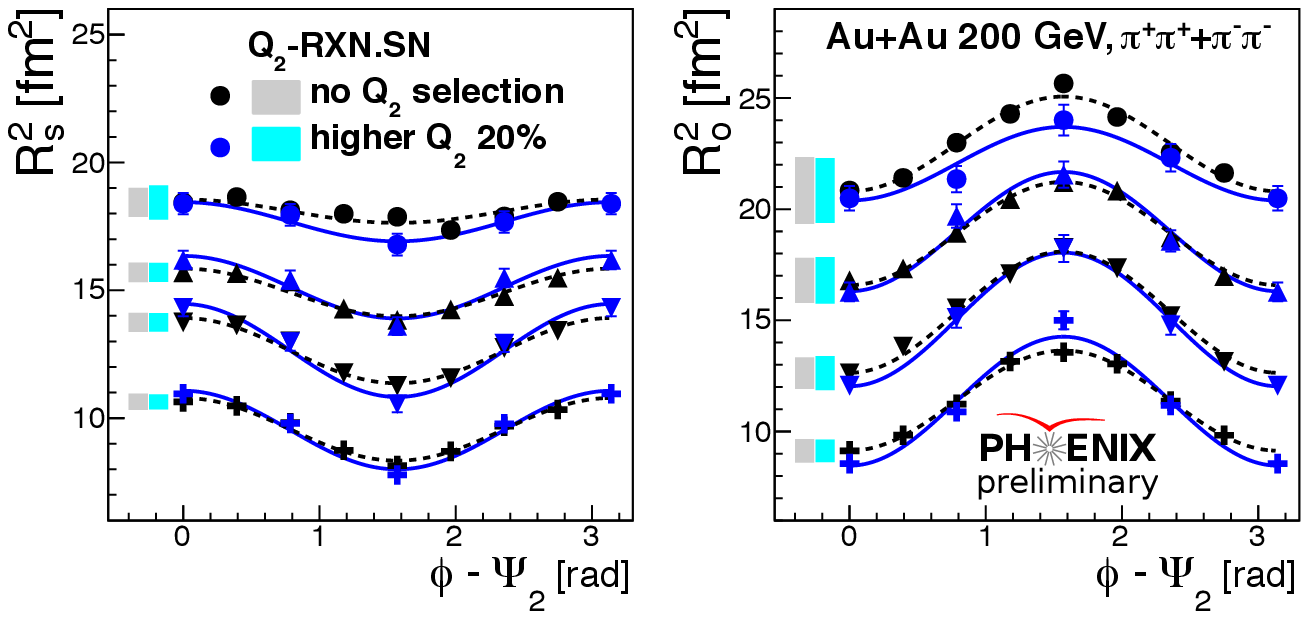}
\end{minipage}\hspace{2pc}%
\begin{minipage}{0.4\textwidth}
\includegraphics[width=0.90\textwidth,clip,trim=30 20 -25 0]{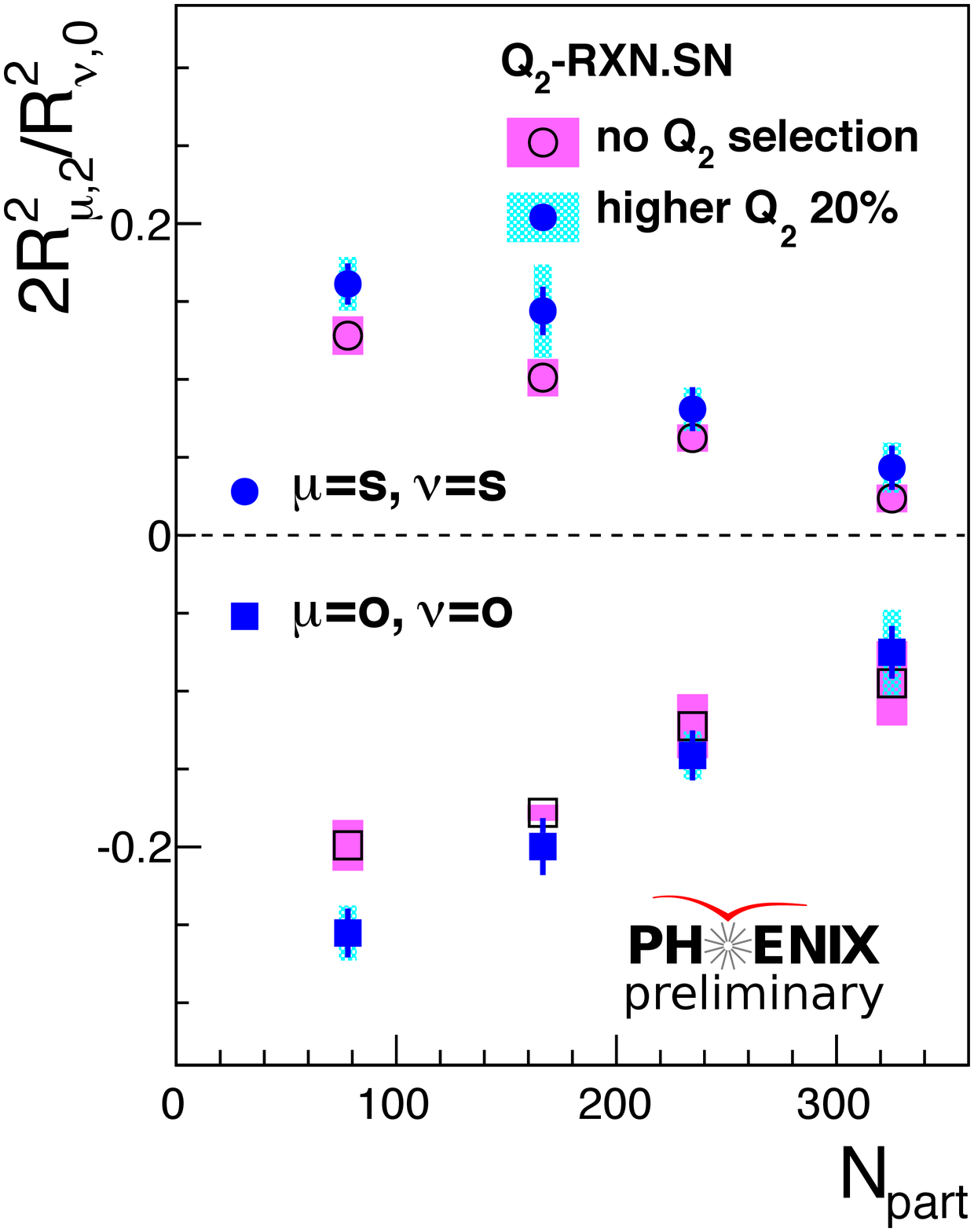}
\end{minipage} 
\caption{\label{fig2} (Left and center) Azimuthal angle dependence of $R_{\rm s}^{2}$ and $R_{\rm o}^{2}$ relative to $\Psi_2$. (Right) 2$^{\rm nd}$-order azimuthal oscillation on $R_{s}^{2}$ and $R_{o}^{2}$. In both panels, results with and without higher $Q_2$ selection are shown.}
\end{figure}

The event shape selection was performed by selecting the magnitude of $2^{\rm nd}$-order flow vectors, $Q_2$,
which were measured by the Reaction Plane Detector (RXN, $1<|\eta|<2.8$).
The $Q_2$ is defined as $Q_2 = \sqrt{Q_{2,x}^{2}+Q_{2,y}^{2}}/\sqrt{\Sigma w_i}$, 
where $Q_{2,x} = \Sigma w_i \cos(2\phi)$, $Q_{2,y} = \Sigma w_i \sin(2\phi)$, and 
$w_i$ reflects the multiplicity within the element $i$ of the RXN.
Figure~\ref{fig1}(left) shows $Q_2$ distributions for two centrality bins, where the higher 20\% and the lower 30\% $Q_2$ events are shown filled.
%mention Bessel-Gaussian?
We have tested the effect of the $Q_2$ selection on charged hadron $v_2$ as shown in Fig.~\ref{fig1}(right), where the $v_2$ is measured at mid-rapidity ($|\eta|<0.35$) and the systematic uncertainty from different event planes of south, north, and both combined RXN is included. It is confirmed that the $Q_2$ selection enhances or decreases the strength of $v_2$.
%
%Fig1------------------------------------------------------------------------------
Then we have applied this technique to the HBT measurement using charged pion pairs. 
Charged pions were identified by the electromagnetic calorimeter (EMCal, $|\eta|<0.35$) and the pair selection cuts at the drift chamber and EMCal were applied to remove the effects of mis-reconstructed tracks and detector inefficiency. For the HBT analysis, pion pairs were analyzed with the out-side-long parameterization~\cite{OSL1,OSL2} in the longitudinally co-moving system. The effect of the event plane resolution was also corrected for both cases with and without $Q_2$ selection.
Figure~\ref{fig2}(left) shows the extracted pion HBT radii, $R_{\rm s}$ and $R_{\rm o}$, as a function of azimuthal pair angle $\phi$ relative to the second-order event plane $\Psi_2$. Results show that the higher $Q_2$ selection increases the oscillation strength compared to the case without $Q_2$ selection. These oscillations of HBT radii are supposed to be sensitive to the final source eccentricity at freeze-out, $\varepsilon_{\rm final}$. Blast-wave studies suggest that the quantity of $2R_{\rm s,2}^{2}/R_{\rm s,0}$ would be a good probe to $\varepsilon_{\rm final}$ in the limit of $k_{\rm T}=0$, where $k_{\rm T}$ denotes a mean pair transverse momentum. The oscillation amplitudes of $R_{\rm s}^2$ and $R_{\rm o}^2$ in a form of the final eccentricity are plotted as a function of the number of participants calculated by Glauber model in Fig.~\ref{fig2}(right). The higher $Q_2$ selection enhances the measured $\varepsilon_{\rm final}$ as well as $v_2$. 
It could be originating from a larger initial eccentricity, although there should be a contribution from the selected flow itself because the radii modulations also depend on the anisotropy in the momentum space~\cite{bw,psi3}.
%
%Fig2------------------------------------------------------------------------------

\section{HBT measurements with event twist selection in AMPT model}
To study the twisted source in the final state, we used the data of Pb+Pb 2.76 TeV collisions simulated using AMPT model (v2.25 with string melting), where the impact parameter was fixed to $8$ fm. For the HBT study, the interference effect between two identical particles, 
$1+\cos(\Delta {\bf r} \cdot \Delta {\bf p})$, was calculated and weighted to the relative pair momentum distributions. Then the correlation functions were reconstructed by taking a ratio of the distributions with and without the weight. Also, all charged pions were allowed to make a pair with each other including $\pi^+\pi^-$ to increase the statistics (the consistency between results for positive and negative pairs was checked).
The event plane was determined using particles in $4<|\eta|<6$, where particles were divided into two sub-groups.
A set of forward and backward event planes ($\Psi_2^{\rm F}, \Psi_2^{\rm B}$) were used for the event cut which requires finite difference between $\Psi_2^{\rm F}$ and $\Psi_2^{\rm B}$ as a event twist selection, whereas the other set of events was used for a reference angle of azimuthal HBT measurement.

\begin{figure}[b]
\includegraphics[width=0.99\textwidth,trim=20 50 20 570]{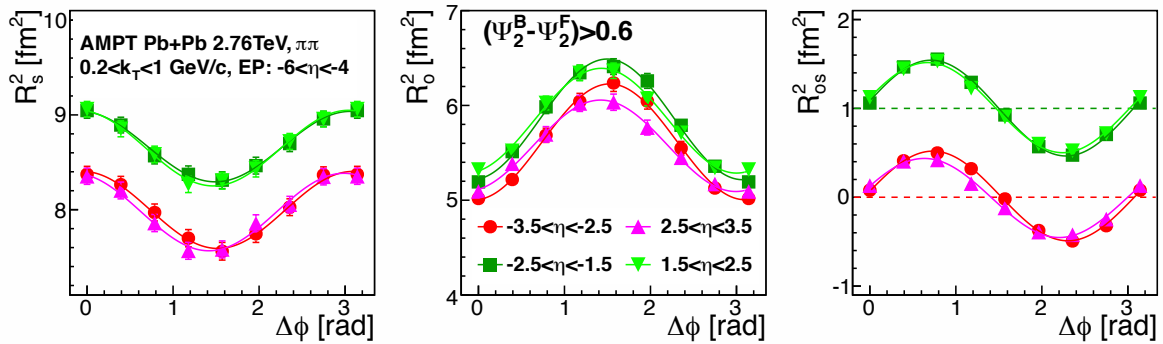}
\caption{\label{fig3} Azimuthal angle dependence of $R_{\rm s}^{2}$, $R_{\rm o}^{2}$, and $R_{os}^2$ relative to the backward $\Psi_2$ with the event cut of $(\Psi_2^{\rm B}-\Psi_2^{\rm F})>0.6$, where the dashed lines show $R_{\rm os}^2=0$.}
\end{figure}

\begin{figure}[t]
\includegraphics[width=0.99\textwidth,clip,trim=0 40 0 560]{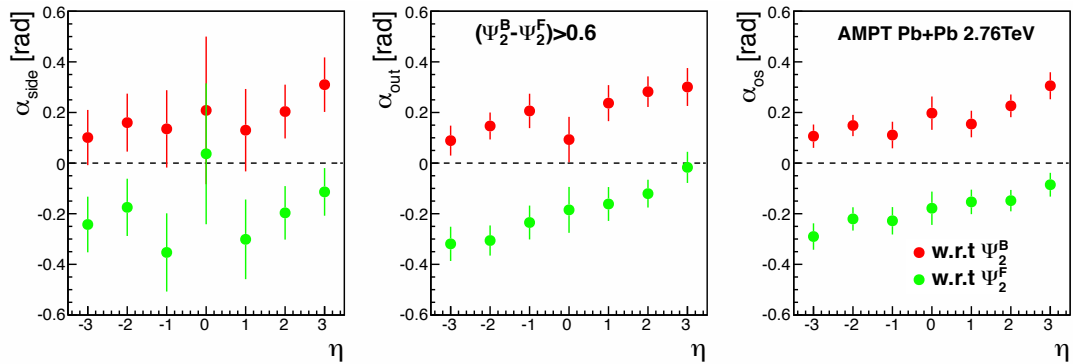}
\caption{\label{fig4} Phase shift parameters $\alpha$ obtained from the $R_{\rm s}^2$, $R_{\rm o}^2$, and $R_{\rm os}$ as a function of $\eta$. Results measured with respect to forward and backward event planes ($\Psi_2^{\rm F}$, $\Psi_2^{\rm B}$) are shown, with event cut of $(\Psi_2^{\rm B}-\Psi_2^{\rm F})>0.6$.}
\end{figure}

Figure~\ref{fig3}(left) shows $R_{\rm s}^2$, $R_{\rm o}^2$, and $R_{\rm os}^2$ as a function of azimuthal pair angle relative to the $\Psi_2^{\rm B}$, $\Delta\phi$, for four $\eta$ regions, where $(\Psi_2^{\rm B}-\Psi_2^{\rm F})>0.6$ was required. The oscillations of three HBT radii measured in positive $\eta$ regions are shifted to negative direction, which is the direction of $\Psi_2^{\rm F}$ in the current event cut. This phase shift can be understood to be a possible twist effect in the final source distribution.
%Fig. 3
%--------------------------------------------------------
%--------------------------------------------------------
These oscillations were fitted with the following functions:
\begin{eqnarray}
R_{\mu}^2 (\Delta\phi) &=& R_{\mu,0}^2 + 2R_{\mu,2}^2 \cos(2 \Delta\phi + \alpha) \,\; ({\rm for} \; \mu=o, s),\\
R_{\mu}^2 (\Delta\phi) &=& R_{\mu,0}^2 + 2R_{\mu,2}^2 \sin( 2\Delta\phi + \alpha) \;\; ({\rm for} \; \mu=os),
\end{eqnarray}
to extract the magnitude of the phase shift, which is taken into account with $\alpha$. The phase shift parameter $\alpha$ obtained from the results with respect to  $\Psi_2^{\rm F}$ and $\Psi_2^{\rm B}$ is plotted as a function of $\eta$ in Fig.~\ref{fig4}. The $\alpha$ increases with going from backward to forward angle in all cases. The variation of $\alpha$ in the $\eta$ dependence is comparable to the difference between results relative to $\Psi_2^{\rm F}$ and $\Psi_2^{\rm B}$ at the same $\eta$. These results indicate that the source in the final state is also twisted due to longitudinal fluctuations in the initial state, as well as the twisted event plane and flow as discussed in Ref.~\cite{ETW1,ETW2}.
%Fig. 4
%--------------------------------------------------------
%--------------------------------------------------------

\section{Summary}
We presented the results of HBT measurements using event shape engineering for collisions recorded with the PHENIX experiment. We found that the higher $Q_2$ selection enhances the measured final source eccentricity as well as $v_2$. Although the model comparison is needed to disentangle both spatial and dynamical effects on the HBT radii, this study clarifies the relation between initial and final 
eccentricity and constrains better the system dynamics.

We have also studied the event twist effect using the AMPT model. When selecting events with finite difference between forward and backward event plane angles, the oscillations of HBT radii are shifted in the phase and the phase shift increases with $\eta$. The results indicate a possible twisted source in the final state preserving the initial twist due to the longitudinal fluctuations. This effect could be measured in experiments at RHIC and the LHC.

Both techniques could be useful to probe and control initial fluctuations in transverse plane and longitudinal directions, 
as well as to study the response of the system to the space-time evolution.

\end{document}